# DETERMINATION OF PHYSICAL AND MECHANICAL PROPERTIES OF SUGARCANE SINGLE-BUD BILLETS

Meimei Wang [1] Qingting Liu[2] Yinggang Ou[2] Xiaoping Zou[2]

（1. Department of Mechanical Engineering, Anyang Institute of Technology, Anyang 455000, China；2. College of Engineering, South China Agricultural University, Wushan Road, Tianhe district, Guangzhou 510642, China）

**Highlights**
- Coefficients of restitution between billet and steel decrease with an increase in drop height and moisture content.
- Coefficients of restitution between billets increase with an increase in moisture content.
- Static friction coefficient between billet and steel and static friction coefficient between billets decrease with an increase in contact area.
- Rolling friction coefficient between billet and steel increase with an increase in angle and interval.
- Rolling fiction coefficient between billets increase first and then decrease with an increase in angle.

*Abstract Determining the physical and mechanical properties of sugarcane single-bud billets is a critical step in the mechanical structure design of a sugarcane planter. In this study, the Tai Tang F66 cultivar sugarcane samples are analyzed. The moisture content of the billets is found to range from 63.78% to 77.72%, and the average density is 244.67 kg/m3. The coefficient of restitution (CoR) of the samples is determined by conducting a drop test wherein the samples are dropped onto a steel plate from different heights. The static friction coefficient (SFC) of four types of samples is determined by the inclined plate method at two orientations. In addition, the rolling friction coefficient (RFC) is determined at three plate inclination angles and sample displacement. The experiment results show that with increasing drop height and moisture content, the billet–steel CoR decreases from 0.625 to 0.458, while the billet–billet CoR increases from 0.603 to 0.698. With an increase in contact area, the billet–steel SFC decreases from 0.515 to 0.377 and the billet–billet SFC decreases from 0.498 to 0.323. With increasing angle and sample displacement, the billet–steel RFC increases from 0.0315 to 0.2175 and the billet–billet RFC increases from 0.0203 to 0.1007. These parameters are useful in the design and optimization of sugarcane single-bud billet planters using EDEM simulation.*

*Keywords. Single-bud billet; Coefficient of restitution; Static friction coefficient; Rolling friction coefficient*

# INTRODUCTION

In sugarcane cultivation, stalks of sugarcane, called billets, are planted. Compared with multi-bud billets, single-bud billets are less bulky, easy to transport, and more economical and have better seed quality (Galal, 2016). Hence, sugarcane planting with single-bud billets has higher production efficiency and lower cost of planting material and is more conducive to automation. However, the productivity of sugarcane billet planters depends on the rate at which billets are dropped by the metering device (Taghinezhad et al., 2014; Robotham et al., 2002). Moreover, the seeding process is closely related to the movement and stress state of billets. Therefore, it is particularly important to analyze the principle of billet movement for the mechanical parameter design and theoretical research of seed-metering devices. Currently, the discrete element method (DEM) is being widely applied in the field of agricultural engineering to analyze the contact between particles and related mechanical components (Zhou et al., 2020; Tan et al., 2021). For the discrete element method simulation in EDEM software, appropriate parameters such as the density of the agricultural material, coefficient of restitution (CoR), static friction coefficient (SFC), and rolling friction coefficient (RFC) must be obtained (Zhou et al., 2021; Mei et al., 2016; Du et al., 2019; Martin et al., 2003). Therefore, this study aims to accurately determine the physical and mechanical properties of the sugarcane single-bud billet.

The mechanical and physical properties of agricultural materials have been extensively researched. The moisture content and density of different agricultural materials are commonly measured by conventional methods (Lorestani et al., 2012; Huang et al., 2018), whereas the method for measuring the CoR, SFC, and RFC differ according to the research object.

The CoR is defined as the ratio of the relative velocities of two objects before and after they collide (Kuwabara et al., 1987). Two simple tests are most frequently conducted to determine the CoR: a drop test and a low-velocity pendulum-based collision test (Horabik et al., 2017). Feng et al. (2017)

calculated the CoR by dropping potatoes from a specified height to a 45° inclined solid flat plate and measuring the horizontal and vertical displacements. Zhou et al. (2021) conducted the drop test to determine the CoR between maize seeds and an organic glass plate and the single pendulum test to determine the CoR between maize seed particles. High-speed video cameras are usually used to obtain the rebound height of objects after collision. Horabik et al. (2017) used a high-speed camera to determine the CoR of three legumes in a single-drop test at different heights and a repeated-bounce test. Mirror reflection is usually used in the test conducted to determine the CoR to easily observe the rebound height of a collided seed (Wang et al., 2021; Jinwu et al., 2017).

The SFC represents the friction characteristics between a material and a surface and is an important parameter for designing the angle of the rake bar chain, rake bar, and seed box of a sugarcane planter. The simplest and most common method to measure the SFC is the inclined plate method. Researchers attach seeds to the plate in various ways to measure the SFC of agricultural materials. Varnamkhasti et al. (2008) placed a cylinder filled with rice on different surfaces to measure the SFC between rice and the surface material. Zhou et al. (2021) glued maize seed particles onto an organic glass plate. Wiącek et al. (2021) used a tilting table equipped with a sample piece of a steel sheet used for a model silo and a frame filled with wheat to measure the SFC. Flat seed units composed of three spaced kernels were used to measure the angle of external friction of cereal kernels on the steel plate (Kaliniewicz et al., 2018; Kaliniewicz et al., 2021). A similar method was used for grass pea (Sadowska et al., 2013); wheat, chickpea, safflower, rye, soybean, and sunflower seeds (Shafaei et al., 2015); rice (Jouki et al., 2012); and melon seeds (Shieshaa et al., 2007). For some seeds whose SFC cannot be directly measured, simulation can be used. Du et al. (2019) conducted a simulation test to calibrate the simulation parameters and obtain the SFC between broad bean seeds. Zhou et al. (2021) used the DEM simulation calibration method to determine the RFC by analyzing the sensitivity of the RFC between seed particles and between seed particles and the organic glass plate. Furthermore, the SFC and DFC of agricultural materials are related to their moisture content (Shafaei et al., 2016) and orientation relative to the direction of motion (Kaliniewicz et al., 2020).

Some studies on the mechanical properties of the sugarcane stalk (i.e., bending resistance, cutting resistance, penetration resistance, and crushing resistance; Bastian et al., 2014), sugarcane leaf sheath (i.e., longitudinal tensile strength, transverse tensile strength, and punch-and-die shear strength; Mou et al., 2013), and sugarcane leaves (i.e., SFC; Khongthon et al., 2014) have been reported. However, the physical and mechanical properties of the sugarcane single-bud billet, which are different from those of the sugarcane stalk, are rarely researched. The results of such a study are useful for the design and optimization of sugarcane planters for single-bud billets. Hence, this study aims to determine the physical and mechanical properties of the sugarcane single-bud billet, hereinafter referred to as the billet.

## MATERIALS AND METHOD

### SAMPLE PREPARATION

The sugarcanes used in the experiments in this study were collected from the farm of Guangdong Guangken Agricultural Machinery Service Co., Ltd., in the city of Zhanjiang, Guangdong Province, China. Upright stems of the Tai Tang F66 variety sugarcane plants that were disease and pest free were collected. Considering the different moisture content of different parts of sugarcane, billets of a length of 60 mm were cut from the top, middle, and root of the collected sugarcanes; 800 sample billets were measured, and the average diameter and weight of the billets are listed in Table 1.

**Table 1 Statistical data of sugarcane samples**

| Variety | Sample number | Average diameter /mm | Standard deviation /mm | Average weight /kg | Standard deviation/ kg |
|---|---|---|---|---|---|
| Tai tang F66 | 800 | 28.7 | 8.7 | 3.532 | 0.146 |

## MEASUREMENT OF THE MOISTURE CONTENT

The billets were heated at 105 ± 2 °C in an air convection oven until they reached a constant weight. The moisture content of the billets was then determined using the oven drying standard method (AOAC, 1990). The moisture content of billets based on the mass difference of billets before and after drying is calculated as follows:

$$w = \frac{m_2 - m_3}{m_2 - m_1} \times 100\%, \tag{1}$$

where w is the moisture content of billets, m1 is the mass of the container holding the billet, m2 is the total mass of the box and billets before drying, and m3 is the total mass of the box and billets after the last drying. Five repetitions of the measurement were obtained for every treatment.

## MEASUREMENT OF BILLET DENSITY

As the billet is cylindrical, its volume $V$ is calculated by Eq. (2). The diameter of the billet was measured by a digital caliper (precision 0.02 mm). The length of the billet cut by a billet cutting machine is 60 mm.

$$V = \pi r^2 h \tag{2}$$

where r is the radius of the billet and h is the height of the billet. The billet density $\rho$ is calculated as follows:

$$\rho = \frac{M}{V}, \tag{3}$$

where M is the billet mass.

## MEASUREMENT OF BILLET COR

The drop test was used to determine the CoR between the billet and steel (billet–steel CoR) and the CoR between billets (billet–billet CoR). Based on the principle of kinematics, the relative velocities before and after collision of the billet with a steel contact plate are related to the height of the fall regardless of air resistance. The CoR (e) is then calculated as follows:

$$e = \sqrt{\frac{h}{H}}, \tag{4}$$

where h is the rebound height of the billet and H is the initial height of the billet.

The experimental setup of drop test includes a platform for fixing the sample and a collision plate. The height from the platform to the collision plate can be adjusted. Two types of collision plates were used for the experiments: a steel plate and a billet plate. The billet samples were adhered to form a billet plate, which minimizes the energy loss due to the rolling and pushing of billets. The samples were positioned at the fixing platform at a certain height above the collision plate, with the axis of the sample parallel to the contact plate, such that the sugarcane peel rather than the cut cross section contacts the plate during collision. Under gravity, the samples were allowed to fall vertically and collide with the steel plate and billet plate from heights of 300 mm and 500 mm. A ruler is fixed on the vertical surface of the collision plane, and a high-speed camera (CR600×2, Optronis, Germany) was used to record the process of falling, colliding, and bouncing of the sample. The highest position at which the sample bounced after impact with the collision plate was marked while replaying the recorded video; the CoR was then analyzed and calculated.

*Measurement of billet SFC*

The SFC was determined using the inclined plate method and the following calculation formula:

$$f_{s=tan\theta}, \tag{5}$$

The static friction angle θ is measured by placing the sample on a steel plate that is horizontally adjusted to gradually increase the angle of inclination until the sample begins to slip (Kaliniewicz et al., 2013).

Fig. 1 shows the experimental device and setup. Four types of test samples, namely one billet, two billets, three billets, and sugarcane peel, were placed horizontally (Fig. 1a) and vertically (Fig. 1b) on the experimental device. To prevent the horizontally placed sample from rolling, only one billet sample was tested vertically. To prevent relative movement and rolling of the sample, the side of the sample of two billets or three billets was fixed. As shown in Fig. 1c, sugarcane peels were flattened and adhered to the square steel plate of the device, allowing the sample to slide down smoothly. This sugarcane peel plate was used to conduct the test for determining the billet–billet SFC.

The procedure for calculating the SFC is as follows. First, four types of samples were placed on the steel and sugarcane peel plates, as shown in Fig. 1. The plate was then gradually lifted until the sample slid down the plate. The digital angle ruler measures the angle of the plate with an accuracy of 0.05°. The experiment was repeated ten times in each group. Finally, the moisture content of samples (except sugarcane peel samples) were measured and recorded.

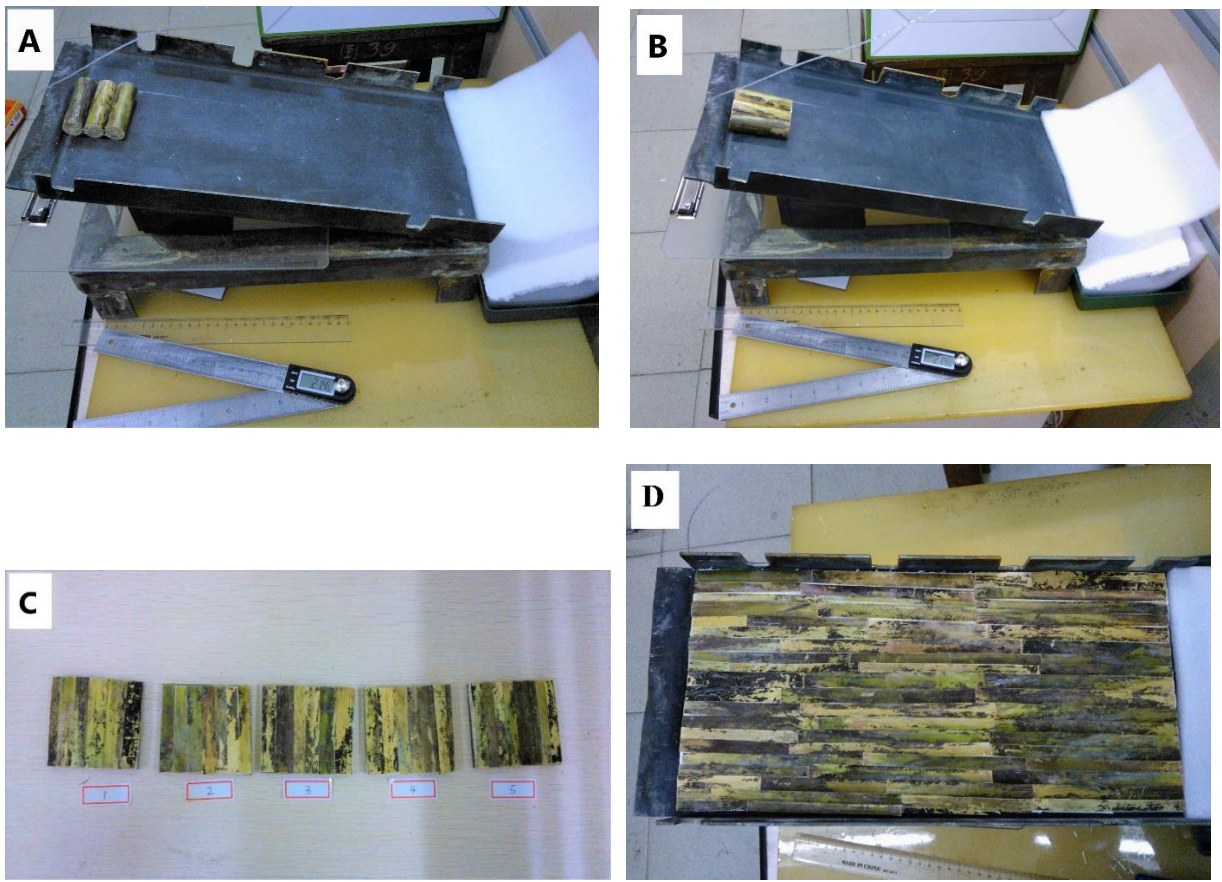

**Fig. 1. Experimental device and samples, A) Three billet samples placed horizontally, B) Three billet samples placed vertically, C) Sugarcane peel samples, D) Sugarcane peel plate.**

*Measurement of billet RFC*

Based on the previous calculation method (Cui et al., 2013; Shan et al., 2009; Zhu et al., 2006) and the billet characteristics, the billet–steel and billet–billet RFCs were calculated by Newton's second law. As shown in Fig. 2, the billet moves uniformly with an initial velocity of zero on the inclined plane, ignoring air resistance and only considering the action of gravity G, supporting force N, and rolling friction force F; then, the following formula is used to calculate the RFC:

$$Ma = Mg\sin\varphi - f_r Mg\cos\varphi, \tag{6}$$

$$L = \frac{1}{2}at^2, \tag{7}$$

$$f_r = \tan\varphi - \frac{2L}{gt^2\cos\varphi}, \tag{8}$$

where fr is the RFC, M is the mass of the billet, a is acceleration, g is the acceleration of gravity, φ is the inclination angle of the inclined plane, L is the displacement of the rolling billet, and t is the time required for displacement L.

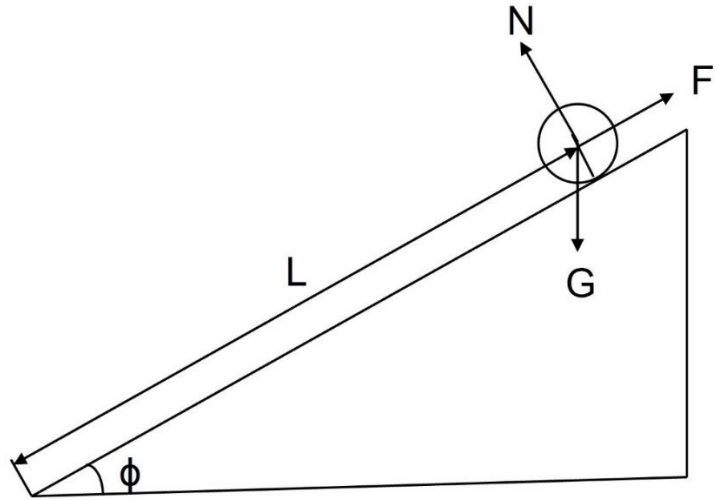

**Fig. 2 Rolling friction force diagram**

The experimental device and setup are shown in Fig. 3. Two photoelectric sensors were attached to the side of the device, and their positions could be adjusted to 10 cm, 20 cm, and 30 cm from sensor 1. The timer was started when the billet passed the position of sensor 1 and ended when it passed the position of sensor 2. To ensure that the billet rolls down smoothly, the angle of the inclined plate was set at three angles: 15°, 30°, and 45°. The experiment was repeated ten times for each group.

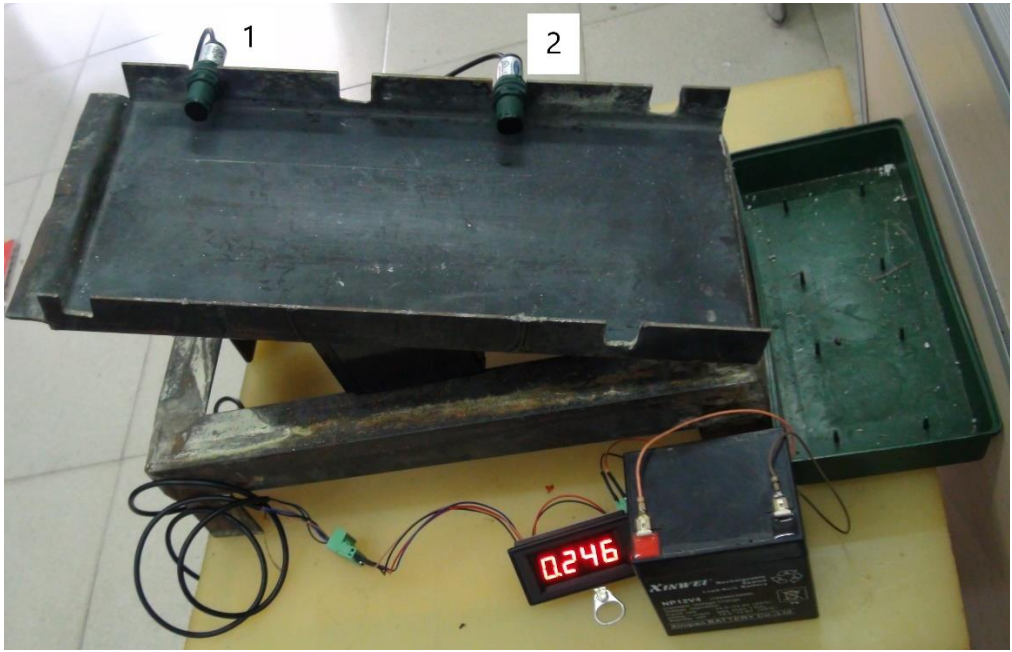

**Fig. 3 RFC measuring system setup**

*Statistical Analysis*

The moisture content and density of the samples were measured using 88 replications and the average values are reported. The statistical parameters, namely the average, minimum, and maximum values, and standard deviations, were computed using Microsoft Excel (2016). The effects of relevant factors on the CoR, SFC, and RFC were analyzed using an independent sample test and one-way analysis of variance (ANOVA) in IBM SPSS Statistics 27 at a significance level of $\alpha = 0.05$.

## RESULTS AND DISCUSSION

### ANALYSIS OF PHYSICAL PROPERTIES RESULTS

Table 2 shows the results of the moisture content and density of 88 samples. Because the samples were collected from different parts of the sugarcane, the moisture content ranged from 63.78% to 77.72%, with an average of 72.74% and a standard deviation of 0.03; the average density was 244.67 kg/m3 with a standard deviation of 10.64 kg/m3.

**Table 2 Moisture content and density of billet samples**

| Sample number | Moisture content % | | | | Density kg/m³ | |
|---|---|---|---|---|---|---|
| | Minimum | Maximum | Mean | Standard deviation | Mean | Standard deviation |
| 88 | 63.78% | 77.72% | 72.74% | 0.03 | 244.67 | 10.64 |

## ANALYSIS OF RESULTS OF BILLET-STEEL AND BILLET-BILLET CoR

The calculated billet–steel CoR is shown in Table 3. It ranged from 0.458 (moisture content 75%) to 0.554 (moisture content 68%) at a drop height of 500 mm and from 0.534 (moisture content 70%) to 0.625 (moisture content 69%) at a drop height of 300 mm. The standard deviation and the standard error were less than 0.0654 and 0.0249, respectively. The differences in the billet–steel CoR at the two drop heights were significant by the independent sample test (P = 0.046 < 0.05); the billet–steel CoR at a drop height of 300 mm was greater than that at 500 mm. This result is similar to Feng et al.'s (2017) observation in case of potatoes: the billet–steel CoR decreased with the drop height. The kinetic energy, deformation, and energy loss increased as the drop height increased, resulting in a decrease in the billet–steel CoR.

**Table 3 Results of CoR between the billet and the steel plate**

| Sample | | Drop height 500 mm | | | Sample | | Drop height 300 mm | | |
|---|---|---|---|---|---|---|---|---|---|
| Moisture content | Sample number | Mean | Standard deviation | Standard error | Moisture content | Sample number | Mean | Standard deviation | Standard error |
| 64% | 5 | 0.553 | 0.0209 | 0.0093 | 69% | 4 | 0.625 | 0.0152 | 0.0076 |
| 68% | 12 | 0.554 | 0.0304 | 0.0088 | 70% | 4 | 0.534 | 0.0271 | 0.0136 |
| 72% | 6 | 0.510 | 0.0219 | 0.0089 | 71% | 7 | 0.563 | 0.0654 | 0.0247 |
| 73% | 9 | 0.503 | 0.0281 | 0.0094 | 72% | 19 | 0.595 | 0.0600 | 0.0138 |
| 74% | 21 | 0.509 | 0.0444 | 0.0097 | 73% | 9 | 0.542 | 0.0424 | 0.0141 |
| 75% | 7 | 0.458 | 0.0161 | 0.0061 | 76% | 5 | 0.537 | 0.0409 | 0.0183 |
| 77% | 7 | 0.516 | 0.0212 | 0.0080 | | | | | |
| Total | 67 | 0.515 | 0.0417 | 0.0051 | Total | 48 | 0.572 | 0.0577 | 0.0083 |

One-way ANOVA revealed that differences in the billet–steel CoR at different moisture contents were significant (P = 0 < 0.05). As shown in and Fig.4 and Fig.5, the billet–steel CoR decreased with increasing moisture content; this observation is consistent with the result for the rice grain (Wang et al., 2021). The quadratic regression equation of moisture content and the billet–steel CoR was obtained. The coefficient of determination (R2) was 0.997 (drop height of 500 mm) and 0.991 (drop height of 500 mm), indicating that the quadratic regression equation fit and predicted well. The higher the moisture content of the billet, the more it deformed when it collided with the steel plate, and the higher its viscosity, the more the energy loss during the collision process. Therefore, the normal separation velocity of the billet rebounding after collision decreased, decreasing the billet–steel CoR.

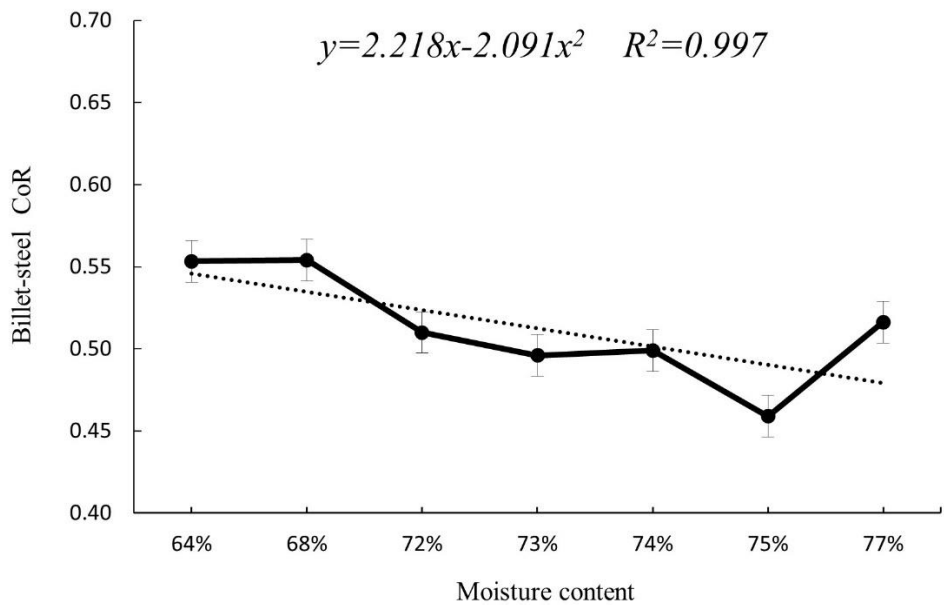

**Fig. 4. Effect of moisture content on the billet-steel CoR at a drop height of 500 mm**

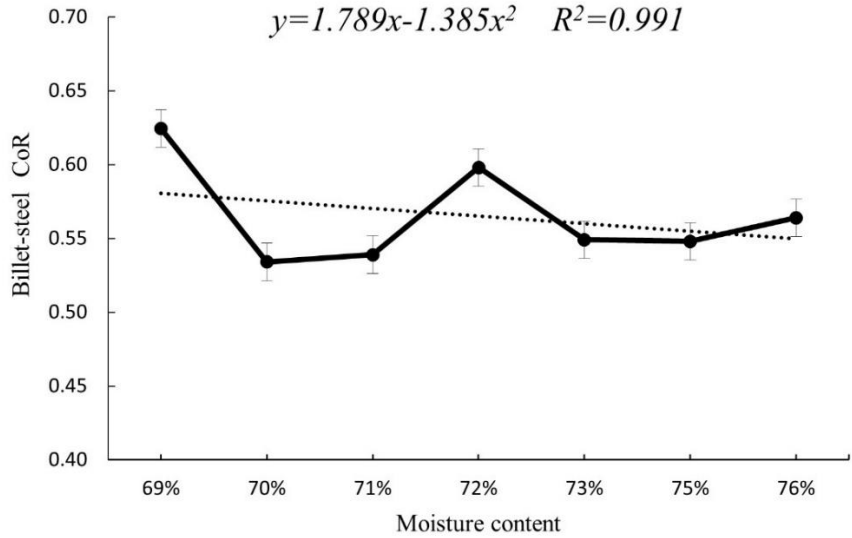

**Fig. 5. Effect of moisture content on the billet-steel CoR at a drop height of 300 mm**

The results of the billet–billet CoR are shown in Table 4. At a drop height of 500 mm, it ranged from 0.634 (moisture content 69%) to 0.698 (moisture content 72%), and at a drop height of 300 mm, it ranged from 0.603 (moisture content 69%) to 0.651 (moisture content 72%). The standard deviation and the standard error were less than 0.0632 and 0.0425, respectively. The differences in the billet–billet CoR at the two drop heights were not significant by the independent sample test ($P = 0.216 > 0.05$).

**Table 4 Results of CoR between two billets**

| Sample | | Drop height 500 mm | | | Sample | Drop height 300 mm | | |
| --- | --- | --- | --- | --- | --- | --- | --- | --- |
| Moisture content | Sample number | Mean | Standard deviation | Standard error | Sample number | Mean | Standard deviation | Standard error |
| 69% | 4 | 0.634 | 0.0520 | 0.0260 | 2 | 0.603 | 0.0601 | 0.0425 |
| 70% | 3 | 0.653 | 0.0445 | 0.0257 | 2 | 0.644 | 0.0127 | 0.0090 |
| 71% | 12 | 0.667 | 0.0532 | 0.0154 | 11 | 0.647 | 0.0632 | 0.0190 |
| 72% | 30 | 0.698 | 0.0223 | 0.0041 | 16 | 0.651 | 0.0352 | 0.0088 |
| 73% | 10 | 0.688 | 0.0348 | 0.0110 | 5 | 0.648 | 0.0247 | 0.0110 |
| 75% | 3 | 0.680 | 0.0170 | 0.0098 | 3 | 0.646 | 0.0422 | 0.0243 |
| 76% | 3 | 0.666 | 0.0286 | 0.0165 | 1 | 0.622 | . | . |
| Total | 65 | 0.683 | 0.0384 | 0.0048 | 40 | 0.646 | 0.0433 | 0.0068 |

Fig. 6 shows that the billet–billet CoR at a drop height of 500 mm was slightly higher than that at a drop height of 300 mm, and they followed a similar trend. One-way ANOVA shows that the differences in the billet–billet CoR on moisture content were significant ($P = 0 < 0.05$). As shown in Fig. 6, the billet–billet CoR increased with increasing moisture content at different drop heights; this trend differs from that of the billet–steel CoR. Meanwhile, overall, the billet–billet CoR was higher than the billet–steel CoR. This is consistent with the results for maize seeds (Zhou et al., 2021) but not with those for rice grains (Wang et al., 2021) or garlic seeds (Guo et al., 2021). This is because the small size of rice and garlic seeds possibly leads to the collision of the seed on the seed plate with a gap rather a complete contact with seed, resulting in a lower bounce height. Zhou et al. (2021) used a single maize seed to collide with the seed, ensuring that the seeds made complete contact with each other. Furthermore, the CoR is affected by factors such as material curvature radii (Wang et al., 2015), friction, and material properties (Ahmad et al., 2016). Because of the larger radius of the billet, the contact area between the billets is smaller than that between the billet and the steel plate, resulting in even smaller deformation. Because of the smooth surface of the billet, when two billets collide completely, the billet loses less energy. Sugarcane is a unidirectional composite biomaterial, and the complexity of the biomaterial may result in a different trend of the CoR, which will require further investigation.

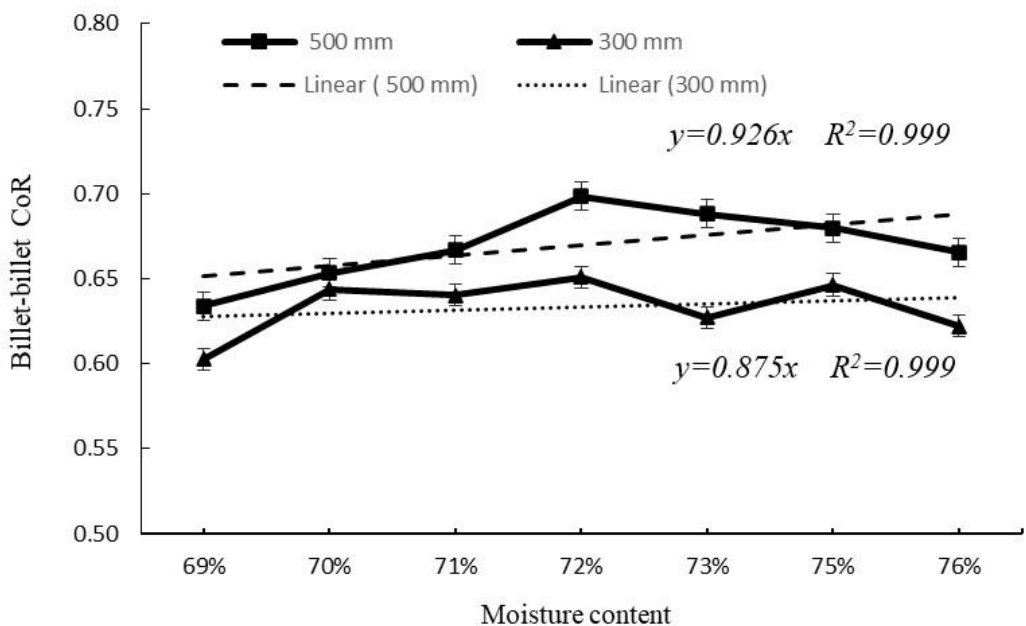

**Fig. 6. Effect of moisture content on the billet-billet CoR.**

**ANALYSIS OF RESULTS OF BILLET-STEEL AND BILLET-BILLET SFC**

According to one-way ANOVA, the effect of moisture content on the billet–steel SFC was not significant in different samples (one billet: P = 0.721 > 0.05; two billets: P = 0.223 > 0.05; three billets: P = 0.604 > 0.05). This can be attributed to the limited moisture content range of mature sugarcane billets. According to the independent sample test, the effect of sample orientation on the billet–steel SFC was also not significant in different samples (two billets: P = 0.534 > 0.05; three billets: P = 0.184 > 0.05; sugarcane peel sample: P = 0.395 > 0.05). However, one-way ANOVA revealed that the effect of different samples on the billet–steel SFC was significant (P = 0 < 0.05). As shown in Fig. 7, the billet–steel SFC of different samples in different orientations varied slightly with increasing moisture content. The billet–steel SFC of one billet sample was generally higher than that of other samples, except for three horizontal billets.

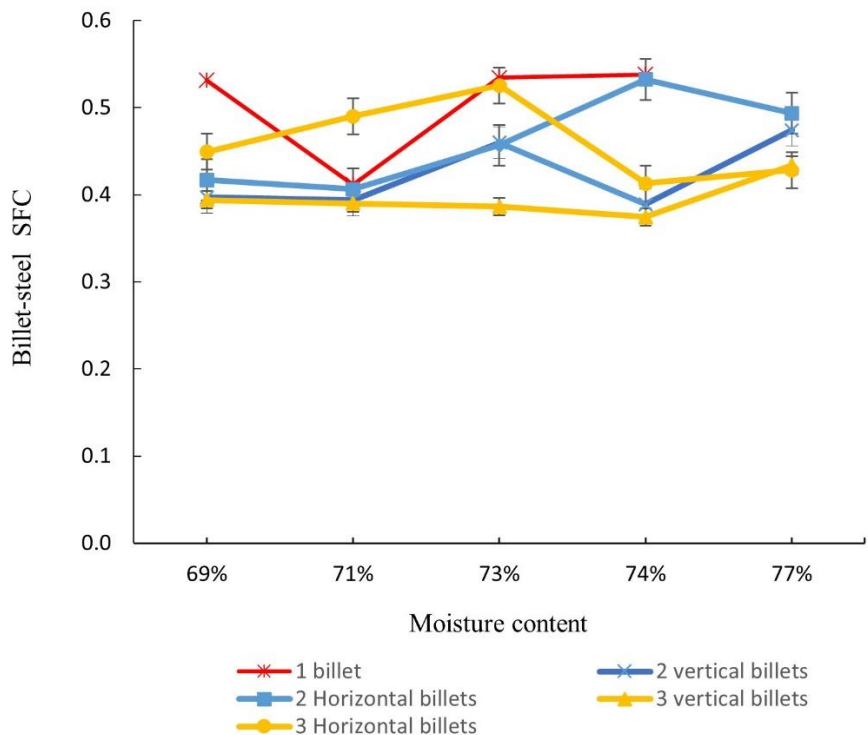

**Fig. 7. Effect of moisture content on the billet-steel SFC.**

Tables 5 and 6 show the billet–steel SFC results of billet samples and sugarcane peel samples,

respectively. The mean of the billet–steel SFC of one billet sample is the highest (mean: 0.515, SD: 0.0068) and that of the sugarcane peel sample is the lowest (mean: 0.377, SD: 0.055). The mean of different samples shows that the billet–steel SFC decreased gradually as the number of billets or contact area increased.

**Table 5. Billet-steel SFC Results of billet samples**

| Moisture content | one billet | | 2 vertical billets | | 2 horizontal billets | | 3 vertical billets | | 3 horizontal billets | |
|---|---|---|---|---|---|---|---|---|---|---|
| | Mean | Standard deviation | Mean | Standard deviation | Mean | Standard deviation | Mean | Standard deviation | Mean | Standard deviation |
| 69% | 0.532 | 0.0149 | 0.397 | 0.0114 | 0.417 | 0.0170 | 0.394 | 0.0099 | 0.449 | 0.0116 |
| 70% | 0.504 | 0.0149 | | | | | | | | |
| 71% | 0.412 | 0.0132 | 0.394 | 0.0061 | 0.406 | 0.0116 | 0.390 | 0.0106 | 0.490 | 0.0039 |
| 72% | 0.514 | 0.0139 | | | | | | | | |
| 73% | 0.535 | 0.0115 | 0.460 | 0.0177 | 0.457 | 0.0090 | 0.386 | 0.0037 | 0.525 | 0.0160 |
| 74% | 0.539 | 0.0129 | 0.389 | 0.0084 | 0.532 | 0.0231 | 0.374 | 0.0040 | 0.413 | 0.0082 |
| 75% | 0.436 | 0.0174 | | | | | | | | |
| 77% | | | 0.474 | 0.0087 | 0.494 | 0.0107 | 0.434 | 0.0099 | 0.428 | 0.0101 |
| Total | 0.515 | 0.0068 | 0.423 | 0.0068 | 0.461 | 0.0088 | 0.396 | 0.0041 | 0.461 | 0.0075 |

**Table 6. Billet-steel SFC Results of sugarcane peel samples**

| Overall sample | | | | Vertical sample | | | | Horizontal sample | | |
|---|---|---|---|---|---|---|---|---|---|---|
| Number | Mean | Standard deviation | Mean | Number | Mean | Standard deviation | Mean | Number | Mean | Standard deviation |
| 10 | 0.377 | 0.055 | | 5 | 0.368 | 0.055 | | 5 | 0.385 | 0.062 |

According to one-way ANOVA, the effect of moisture content on the billet–billet SFC was not significant in different samples (one billet: $P = 0.311 > 0.05$; two billets: $P = 0.094 > 0.05$; three billets: $P = 0.196 > 0.05$). By the independent sample test, the effect of sample orientation on the billet–billet SFC was also not significant in different samples (two billets: $P = 0.11 > 0.05$; three billets: $P = 0.571 > 0.05$; sugarcane peel sample: $P = 0.991 > 0.05$). However, one-way ANOVA revealed that the effect of different samples on the billet–billet SFC was significant ($P = 0 < 0.05$). As shown in Fig. 8, the billet–billet SFC of different samples in different orientations varied slightly with increasing moisture content. The billet–billet SFC of one billet sample was generally greater than that of other samples.

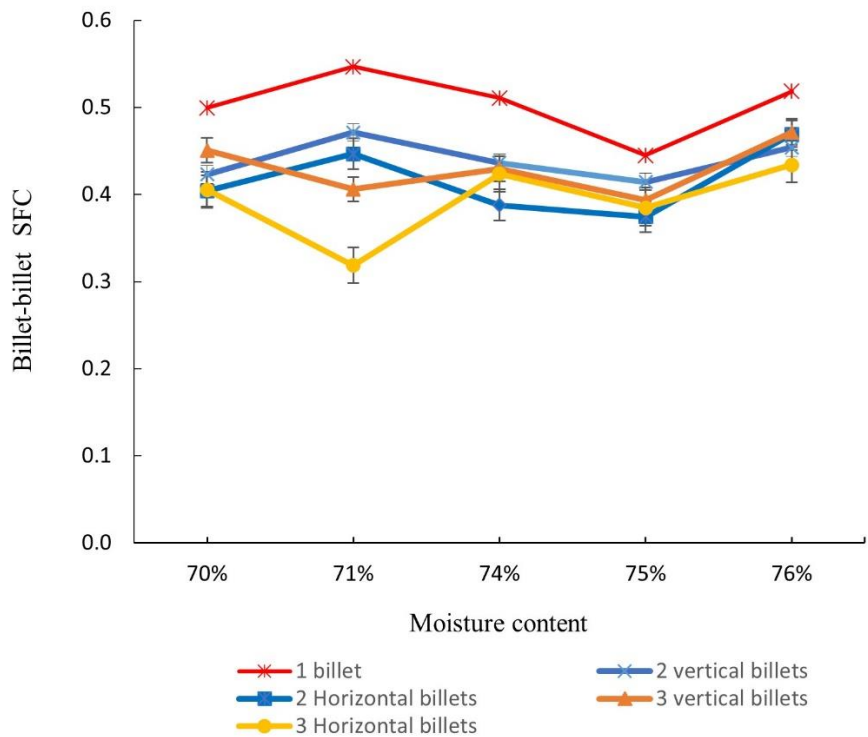

**Fig. 8. Effect of moisture content on the billet-billet SFC.**

Tables 7 and 8 show the billet–billet SFC results of billet samples and sugarcane peel samples, respectively. The mean billet–billet SFC of one billet sample is the highest (mean: 0.498, SD: 0.0059) and that of the sugarcane peel sample is the lowest (mean: 0.323, SD: 0.044). Similarly, to the billet–steel SFC, the billet–billet SFC decreased with increasing contact area.

**Table 7. Billet-billet SFC results of billet samples**

| Moisture content | one billet | | 2 vertical billets | | 2 horizontal billets | | 3 vertical billets | | 3 horizontal billets | |
|---|---|---|---|---|---|---|---|---|---|---|
| | Mean | Standard deviation | Mean | Standard deviation | Mean | Moisture content | Mean | Standard deviation | Mean | Standard deviation |
| 70% | 0.500 | 0.0129 | 0.423 | 0.0097 | 0.404 | 0.0317 | 0.451 | 0.0108 | 0.405 | 0.0138 |
| 71% | 0.547 | 0.0151 | 0.471 | 0.0090 | 0.447 | 0.0099 | 0.406 | 0.0102 | 0.319 | 0.0033 |
| 72% | 0.502 | 0.0134 | | | | | | | | |
| 74% | 0.511 | 0.0156 | 0.436 | 0.0123 | 0.388 | 0.0063 | 0.429 | 0.0059 | 0.423 | 0.0077 |
| 75% | 0.445 | 0.0111 | 0.414 | 0.0131 | 0.374 | 0.0058 | 0.393 | 0.0100 | 0.385 | 0.0130 |
| 76% | 0.518 | 0.0109 | 0.454 | 0.0104 | 0.469 | 0.0108 | 0.471 | 0.0071 | 0.434 | 0.0044 |
| 77% | 0.542 | 0.0171 | | | | | | | | |
| 78% | 0.423 | 0.0194 | | | | | | | | |
| Total | 0.498 | 0.0059 | 0.440 | 0.0056 | 0.416 | 0.0082 | 0.430 | 0.0054 | 0.393 | 0.0064 |

**Table 8. Billet-billet SFC Results of sugarcane peel samples**

| Overall sample | | | | Vertical sample | | | | Horizontal sample | | |
|---|---|---|---|---|---|---|---|---|---|---|
| Number | Mean | Standard deviation | Mean | Number | Mean | Standard deviation | Mean | Number | Mean | Standard deviation |
| 10 | 0.323 | 0.044 | | 5 | 0.352 | 0.036 | | 5 | 0.294 | 0.032 |

**ANALYSIS OF RESULTS OF BILLET-STEEL AND BILLET-BILLET RFC**

The significance value listed in Table 9 confirms that the moisture content in the 70%–75% range had no significant effect on the billet–steel RFC at different plate inclination angles and sample displacements, except at 30° and 10 cm, respectively. The reason for this difference can be attributed to the fact that the moisture content range of this group of test samples is too narrow to be representative (73%–75%); therefore, the effect of moisture content is no longer considered in the following discussion.

**Table 9. Results of billet-steel rolling friction test and ANOVA results of moisture content**

| Angle | 15° | 15° | 15° | 30° | 30° | 30° | 45° | 45° | 45° |
|---|---|---|---|---|---|---|---|---|---|
| Sample displacement | 10 cm | 20 cm | 30 cm | 10 cm | 20 cm | 30 cm | 10 cm | 20 cm | 30 cm |
| Number | 15 | 15 | 15 | 12 | 15 | 15 | 14 | 15 | 15 |
| Mean | 0.0315 | 0.0580 | 0.0737 | 0.0969 | 0.1105 | 0.1536 | 0.0634 | 0.1567 | 0.2175 |
| Standard deviation | 0.0217 | 0.0168 | 0.0150 | 0.0331 | 0.0348 | 0.0227 | 0.0318 | 0.0277 | 0.0405 |
| Standard error | 0.0056 | 0.0043 | 0.0039 | 0.0096 | 0.0090 | 0.0059 | 0.0085 | 0.0071 | 0.0105 |
| F | 0.325 | 0.597 | 1.036 | 9.287 | 3.45 | 2.052 | 0.618 | 0.485 | 1.316 |
| Sig | 0.855 | 0.673 | 0.436 | 0.006 | 0.051 | 0.163 | 0.661 | 0.747 | 0.329 |

The ANOVA results given in Table 10 indicate that the effect of inclination angle and sample displacement on the billet–steel RFC was significant. According to Fig. 8 and Table 9, the billet–steel RFC exhibited an upward trend as the angle and sample displacement increased. The billet–steel RFC increased from 0.0315 (angle of 15°, sample displacement of 10 cm) to 0.2175 (angle of 45°, sample displacement of 30 cm). Because of the irregular shape of the billet, the speed of the billet sample may increase as the angle increases; consequently, the billet may not only roll but also slide and bounce in the movement process. Hence, the work of rolling friction does not entirely account for the energy loss. The larger the sample displacement, the more likely the billet movement track will deviate from the straight line, resulting in greater energy loss, and, in turn, leading to the greater billet–steel RFC and error.

**Table 10 ANOVA results of inclination angle and sample displacement in the billet–steel rolling friction test**

|     | Angle | | | Sample displacement | | |
| --- | --- | --- | --- | --- | --- | --- |
|     | 10 cm | 20 cm | 30 cm | 15° | 30° | 45° |
| F   | 17.091 | 48.469 | 97.912 | 20.904 | 13.156 | 75.937 |
| Sig | 0 | 0 | 0 | 0 | 0 | 0 |

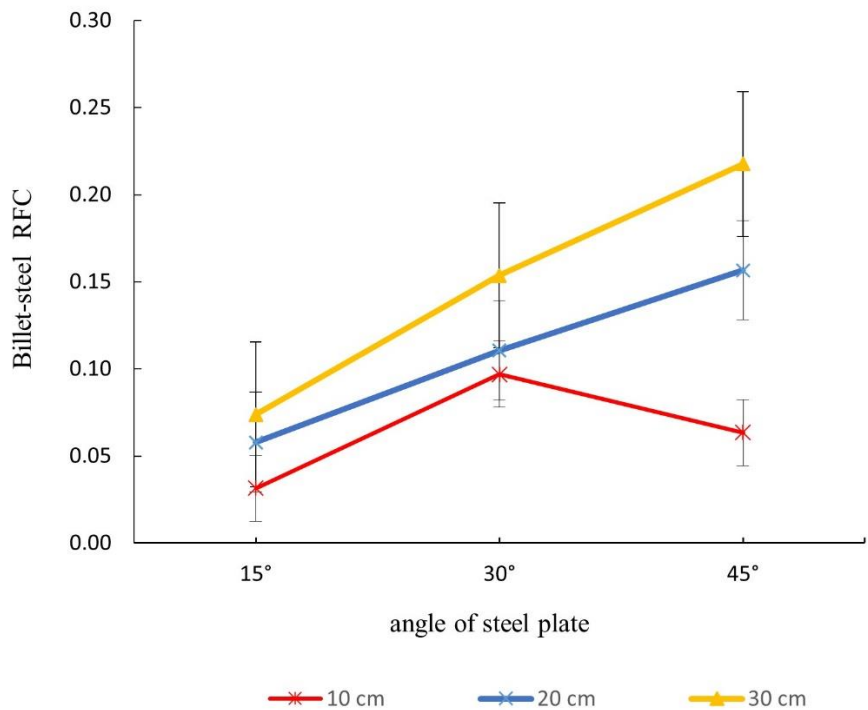

**Fig. 9 Effect of inclination angle and sample displacement on the billet–steel RFC**

The moisture content (69%–78%) had no significant effect on the billet–billet RFC at different angles and sample displacements as shown in Table 11; by contrast, according to one-way ANOVA, the effect of angle and sample displacement on the billet–billet RFC was significant as shown in Table 12.

**Table 11 Results of the billet–billet rolling friction test and ANOVA results of moisture content**

| Angle | 15° | 15° | 15° | 30° | 30° | 30° | 45° | 45° | 45° |
|---|---|---|---|---|---|---|---|---|---|
| Sample displacement | 10 cm | 20 cm | 30 cm | 10 cm | 20 cm | 30 cm | 10 cm | 20 cm | 30 cm |
| Number | 11 | 15 | 15 | 15 | 15 | 15 | 15 | 15 | 13 |
| Mean | 0.0203 | 0.0425 | 0.0621 | 0.0506 | 0.0689 | 0.1007 | 0.0411 | 0.0467 | 0.0687 |
| Standard deviation | 0.0137 | 0.0118 | 0.0112 | 0.0415 | 0.0439 | 0.0351 | 0.0204 | 0.0251 | 0.0351 |
| Standard error | 0.0041 | 0.0030 | 0.0029 | 0.0186 | 0.0113 | 0.0091 | 0.0043 | 0.0036 | 0.0097 |
| F | 0.546 | 1.143 | 0.663 | 119.070 | 0.450 | 0.368 | 1.908 | 0.930 | 0.702 |
| Sig | 0.779 | 0.448 | 0.712 | 0.067 | 0.854 | 0.904 | 0.139 | 0.503 | 0.690 |

**Table 12. ANOVA results of angle and sample displacement in the billet–billet rolling friction test**

|  | Angle | | | Sample displacement | | |
|---|---|---|---|---|---|---|
|  | 10 cm | 20 cm | 30 cm | 15° | 30° | 45° |
| F | 4.613 | 3.527 | 7.48 | 37.825 | 3.945 | 3.489 |
| Sig | 0.021 | 0.042 | 0.002 | 0 | 0.029 | 0.045 |

The billet–billet RFC increased initially and then decreased with increasing angle, as shown in Fig. 10 and Table 11. However, the billet–steel RFC showed a continuously increasing trend as the sample displacement increased. This is possibly because the surface of the sugarcane peel plate is not completely flat, due to which when the plate inclination is high, the sample may not contact the surface for a short duration owing to the beating, resulting in a lower billet–billet RFC. Therefore, the inclined plate should be kept at an appropriate angle. The billet-steel RFC ranged from 0.0203 (angle of 15°, sample displacement of 10 cm) to 0.1007 (angle of 30°, sample displacement of 30 cm).

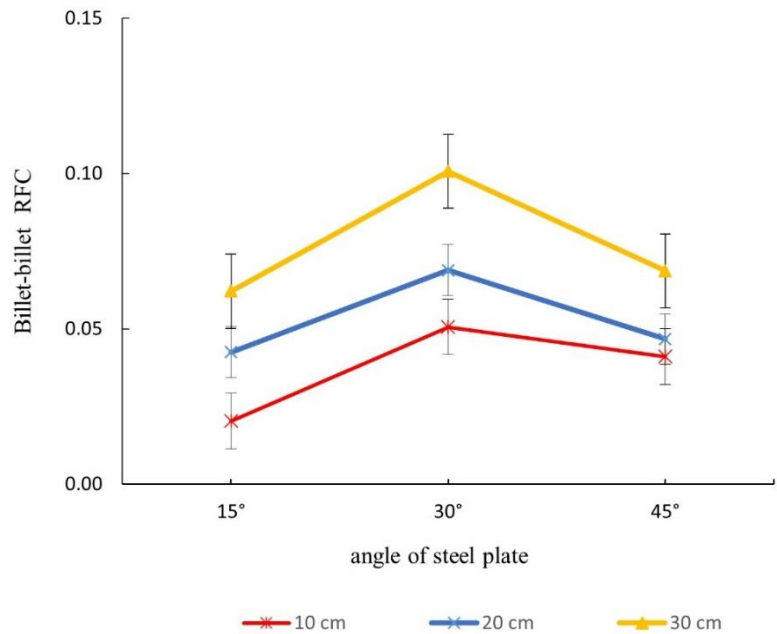

**Fig. 10. Effect of angle and sample displacement on the billet–billet RFC**

These physical and mechanical parameters of sugarcane billets are affected by various factors and have a certain range of variation; therefore, they can serve as input data for EDEM simulation in the sugarcane planter design process. Because the input parameters of the simulation are unique and determined values, they may need to be selected and calibrated in the EDEM simulation test. Furthermore, laboratory experimental results can be used to verify the accuracy of the parameters.

## CONCLUSIONS

In this study, the physical and mechanical parameters of sugarcane billets were measured on the designed experimental device. In addition, the effects of various factors on the parameters were analyzed. The conclusions of the study are as follows:

1. The moisture content of the sugarcane single-bud billet ranged from 63.78% to 77.72% depending on the part of sugarcane tested. The average density was 244.67 kg/m$^3$ with a standard deviation of 10.64 kg/m$^3$.

2. The effect of drop height and moisture content on the billet–steel CoR was significant, with the billet–steel CoR decreasing from 0.625 (moisture content, 69%; drop height, 300 mm) to 0.458 (moisture content, 75%; drop height, 500 mm) as the drop height and moisture content increased. By contrast, the billet–billet CoR was not significantly affected by the drop height; it increased from 0.603 (moisture content, 69%) to 0.698 (moisture content, 72%).

3. The effect of different samples on the billet–steel SFC was significant, while the effects of moisture content and sample orientation were not. With an increase in the contact area, the billet–steel SFC decreased from 0.515 to 0.377. The billet–billet SFC, which decreased from 0.498 to 0.323, followed a similar trend.

4. The plate inclination angle and sample displacement had a significant effect on the billet–steel RFC, while the moisture content did not. With increasing inclination angle and sample displacement, the billet–steel RFC increased from 0.0315 (angle, 15°; sample displacement, 10 cm) to 0.2175 (angle, 45°; sample displacement, 30 cm). The billet–billet RFC followed the same trend as the billet-steel RFC, except with respect to angle. The billet–billet RFC first increased and then decreased as the angle increased; it ranged from 0.0203 (angle, 15°; sample displacement, 10 cm) to 0.1007 (angle, 30°; sample displacement, 30 cm).

5. In this study, the parameters of sugarcane billets are required to be selected and calibrated in the EDEM simulation to verify their accuracy. Moreover, only the Tai Tang F66 cultivar sugarcane samples were analyzed; in the future, the parameters of different sugarcane varieties will be measured and compared.

**ACKNOWLEDGEMENTS**

This research was supported by China Agriculture Research System of MOF and MARA and Guangdong Provincial Team of Technical System Innovation for Sugarcane Sisal Industry (2019KJ104-11).